\documentclass[twocolumn,showpacs,preprintnumbers,superscriptaddress,amsmath,amssymb,nofootinbib]{revtex4-1}

\usepackage{amsfonts}
\usepackage{amsmath}
\usepackage{dcolumn}
\usepackage{epsfig}
\usepackage{graphicx}% Include figure files
\usepackage{subfigure}
\usepackage{dcolumn}% Align table columns on decimal point
\usepackage{bm}% bold math
\usepackage{times}
\usepackage{xcolor}
\usepackage{float}
\usepackage{soul}
\usepackage{url}
\usepackage{color}
\usepackage{hyperref}
\usepackage{graphicx}% http://ctan.org/pkg/graphicx
\usepackage{lipsum}% http://ctan.org/pkg/lipsum
\usepackage{ccaption}%

\begin{document}

\title{Identify Hidden Spreaders of Pandemic over Contact Tracing Networks}

\author{Shuhong Huang}\thanks{These two authors contributed equally}
\affiliation{Institute of Neuroscience, Technical University of Munich, Munich 80802, Germany}
\author{Jiachen Sun}\thanks{These two authors contributed equally}
\affiliation{MIT Center for Collective Intelligence, Cambridge 02142, MA, USA}
\author{Ling Feng}
\affiliation{Institute of High Performance Computing, A*STAR, 138632 Singapore}
\affiliation{Department of Physics, National University of Singapore, Singapore 117551}
\author{Jiarong Xie}
\affiliation{School of Data and Computer Science, Sun Yat-sen University, Guangzhou 510006, China}
\author{Dashun Wang}
\affiliation{Kellogg School of Management, Northwestern University, Evanston, IL, USA}
\author{Yanqing Hu}\email{huyanq@mail.sysu.edu.cn}
\affiliation{School of Data and Computer Science, Sun Yat-sen University, Guangzhou 510006, China}
\date{\today}

\begin{abstract}
\textbf{The COVID-19 infection cases have surged globally, causing devastations to both the society and economy. A key factor contributing to the sustained spreading is the presence of a large number of asymptomatic or hidden spreaders, who mix among the susceptible population without being detected or quarantined. Here we propose an effective non-pharmacological intervention method of detecting the asymptomatic spreaders in contact-tracing networks, and validated it on the empirical COVID-19 spreading network in Singapore.
We find that using pure physical spreading equations, the hidden spreaders of COVID-19 can be identified with remarkable accuracy.
Specifically, based on the unique characteristics of COVID-19 spreading dynamics, we propose a computational framework capturing the transition probabilities among different infectious states in a network, and extend it to an efficient algorithm to identify asymptotic individuals. Our simulation results indicate that a screening method using our prediction
 outperforms machine learning algorithms, e.g. graph neural networks, that are designed as baselines in this work, as well as random screening of infection's closest contacts widely used by China in its early outbreak. Furthermore, our method provides high precision even with incomplete information of the contract-tracing networks. Our work can be of critical importance to the non-pharmacological interventions of COVID-19, especially with increasing adoptions of contact tracing measures using various new technologies. Beyond COVID-19, our framework can  be useful for other epidemic diseases that also feature asymptomatic spreading.}
\end{abstract}
\maketitle

%\section{Introduction}
As the COVID-19 pandemic continues to spread at rapid rates~\cite{hui2020continuing, Megan2020pandemic,world2020director}, and the development of effective pharmacological treatments is still uncertain according to WHO, non-pharmacological interventions like isolation of the infectious through quarantines~\cite{hellewell2020feasibility, maier2020effective} are the most effective and possibly the only means of containing the continued outbreaks, as it effectively reduces the person to person transmissions ~\cite{chan2020familial}. Yet, unlike other infectious diseases like SARS and Ebola,
%COVID-19 is unique that a large portion of its infected population is asymptotic~\cite{gao2020systematic}, with average 19 days of detoxification period~\cite{long2020clinical}, and as many as $13\%$~\cite{kimball2020asymptomatic} do not exhibit and clinic symptoms until self-recovery.
COVID-19 is unique in that a large portion of its infected population is mild or asymptotic~\cite{gao2020systematic}. Even some of the asymptotic infections do not exhibit any clinic symptoms until self-recovery \cite{kimball2020asymptomatic,long2020clinical}.
%with average 19 days of detoxification period~\cite{long2020clinical}, and as many as $13\%$~\cite{kimball2020asymptomatic} do not exhibit and clinic symptoms until self-recovery.
Without being detected and subsequently quarantined, the asymptomatic population {(i.e. hidden spreaders)} sustains the ongoing spreading of the disease to the susceptible population unknowingly~\cite{liu2020locally,rothe2020transmission}. This poses a major challenge in the effective mitigation of the pandemic spreading.
Furthermore, empirical studies have shown that such asymptomatic infections accounts for a large proportion of the population~\cite{quilty2020effectiveness, byambasuren2020estimating,nishiura2020estimation,day2020covid, yu2020covid,mizumoto2020estimating, Heneghan2020asymptomatic}, as much as up to 80\%~\cite{Heneghan2020asymptomatic}.
%Studies have shown that such asymptomatic population could account for 17.9$\%$ to 30.8$\%$~\cite{nishiura2020estimation, mizumoto2020estimating} of the cases, while further studies have indicated that as much as 60\% of the infected cases are not identified as they are either asymptomatic or showing only mild symptoms, ~\cite{qiu2020covert}.
Currently, estimation of the asymptomatic cases is done through exhaustive screening of close contacts of the known infected cases in the contact tracing networks~\cite{mizumoto2020estimating}. This untargeted method requires  large amount of resources and is time consuming, that in turn leads to ineffective or delayed interventions to quarantine the asymptomatic cases.  Hence, a targeted screening in the contact tracing network is pertinent, such that asymptomatic individuals can be estimated with high precision for intervention and spreading mitigation.

Here we incorporate the empirical characteristics of the COVID-19 spreading dynamics into a Markovian  process, i.e. vectors that represent the different infection stages and their associated transition probabilities. By embedding the transition process into a contact tracing network that includes the known infected nodes (individuals), we develop a method that  predicts the infectious states of the rest of the network with high precision. By combining such predictions with the network structure, we then derive the spreading power of every node taking into account of both its infectious state and its specific location in the network, such that screening of the asymptomatic can be prioritised accordingly.
The effectiveness of our method is validated by empirical data from two COVID-19 transmission networks in Singapore. Moreover, in the simulated COVID-19 transmission experiment of contact-tracing network, we find that a screening scheme designed by the proposed computational framework outperforms several machine-learning baselines designed in this work and the random screening of infection neighbors. The latter was widely used in early COVID-19 outbreaks in China.
%We validate our method on two empirical datasets in Singapore, outperforming widely adopted graph neural network algorithms and random screening. The later was widely used in early COVID-19 outbreaks in China.
Furthermore, even in the realistic situation of incomplete information on the contact tracing network, with missing links or sub networks consisting of only contacts of the infected cases, our method retains high accuracy. Thus our method is highly effective in asymptomatic case estimation and can be implemented to any contact-tracing networks either constructed manually~\cite{againstcovid19.com} {or through technological means~\cite{kondylakis2020covid} such as Bluetooth~\cite{drew2020rapid,ferretti2020quantifying}, GPS~\cite{chang2020mobility} and digital check-in check-out technologies~(e.g. health QR codes~\cite{mozur2020coronavirus} widely used in China)}.
		
\section{results}	
%The spread of the infectious disease occurs over the contact network. Such contact-tracing network can be represented as $G(\mathcal{V},\mathcal{E})$, where $\mathcal{V}$ is the node set representing individuals and $\mathcal{E}$ is the link set representing contact relations between individuals. Commonly at a certain time $T$, the set of infected individuals (nodes) that are symptomatic have been identified, and their infection history regarding infection time and recovery time are also observed. Their close contacts during the infectious period are effectively captured in the contact tracing network.

%Since a high proportion of asymptomatic can exist for COVID-19,

Given the spreading of the COVID-19 occurring over the contact network, the challenge is to identify asymptomatic nodes with the information of infected symptomatic individuals (nodes) that have been identified from a certain time $T$. We approach this by estimating the probability of each node being in the infected state as illustrated in Fig.~1. Specifically, we first construct the transition dynamical equations among different infection stages and states based on the empirically observation of COVID-19 disease progression. The set of transition equations is then combined with the contact network topology and data on the observed infection history to deduce the state of each node in the network.
		
As observed in many clinical studies, the {hidden spreaders} of COVID-19 fall into two different categories. One is the \textbf{presymptomatic} infections who are asymptomatic and infected, but will later develop clinical symptoms (e.g. fever, cough, dyspnea, etc.); The other type corresponds to the \textbf{asymptomatic} patients who carry the virus but have never exhibited any symptoms until recovery. As a result, an individual can have a total of 5 different states in the process of COVID-19 spreading~(see Fig.~1a), namely, Susceptible~($S$), Presymptomatic~($P$), Asymptomatic~($A$), Symptomatic Infectious ($I$) and Recovered~($R$). Since the infectious duration in the states of $P$, $I$ and $A$ follows a specific probability distribution, here we further break down the $P, I$ and $A$ states into finer states representing the progression in each of the 3 states, i.e., the number of days passed since the beginning of the states.
For better clarity, we denote $t$ as the number of days of the COVID-19 evolution on the entire network and $d$ as the number of days in a particular infected state for a particular individual. Since an individual $i$ can be at any stage in the process, we can use $\mathbf{Z}_i(t)$ to represent the state probabilities at time $t$:
\begin{equation}
\begin{aligned}
\mathbf{Z}_i(t) = &(S_i(t),P^1_i(t),...,P^d_i(t), ..., I^1_i(t),...,\\& I^d_i(t),...,A^1_i(t),...,A^d_i(t),...,R_i(t))
\end{aligned}
\end{equation}		
		where $S_i(t)$ and $R_i(t)$ is the probability that the individual $i$ is susceptible and recovered at day $t$, respectively. $P^d_i(t), I^d_i(t)$ and $A^d_i(t)$ are the probabilities that $i$ is in the state of $P$, $I$ and $A$ for $d$ days at the time of $t$.
Since all of asymptomatic, presymptomatic and symptomatic states are infected states, their total probability corresponds to that of a node is infectious, and we use $C_i(t)$ to represent it:
\begin{equation}
C_i(t) = P_i(t)+I_i(t)+A_i(t)
\end{equation}
where $P_i(t) = \sum_{d=1}^\infty P^d_i(t)$, $I_i(t) = \sum_{d=1}^\infty I^d_i(t)$, $A_i(t) = \sum_{d=1}^\infty A^d_i(t)$. Throughout this work we use $C_i(t)$ as a key indicator to infer whether an individual is infected .

From here, we can extract the probability transition dynamics among the 5 different states as follows. First, for a node who is in the susceptible state $S$ at $t$, its next state at $t+1$ will be jointly determined by the state of its neighbors in the network at $t$. Specifically, the probability of a node $i$ in $S$ state remains in $S$ on day $t+1$  (i.e., not infected by any of its infected neighbor on the next day) is:
\begin{equation}
S_i(t+1) = S_i(t)\cdot\prod_{j\in \partial i}(1-\mathcal{F}(t,j,\beta))
\end{equation}
where $\partial i$ represents the set of neighbors~(contacts) of $i$ in the network, $\mathcal{F}(t,j,\beta)$ represents the probability that $i$ is infected by $j$. This can only happen if $j$ is in the infected state on day $t$ (probability $C_j(t)$), and happens to transmit it to $i$ (probability $\beta$). Then we have:		
%		the average daily infection capacity of the node $j$, which is equivalent to the probability of node $j$ being infected at day $t$ multiplied by the probability of close contacts of infected patients being infected within one day $\beta$, i.e.:
\begin{equation}
\mathcal{F}(t,j,\beta)=C_j(t)\cdot\beta
\end{equation}		
Here $\beta$ can be estimated from the empirically observed disease reproduction number $R_0$ for COVID-19 and the average number of neighbors in the contact tracing network $\langle k \rangle$. Specificaly,  $\beta = \frac{R_0}{\lambda\langle k \rangle}$, where $\lambda$ is the average time a susceptible person carries the virus, which can be expressed as
$\lambda = p\cdot\mu_{A}+(1-p)\cdot(\exp(\mu_P+\frac{\sigma_P^2}{2})+\mu_I)$,
where $p$ is the proportion of asymptomatic infected cases, $\mu_{A},\exp(\mu_P+\frac{\sigma_P^2}{2}),\mu_{I}$ are the average time of the virus carried by infected individuals in $A$, $P$ and $I$ states \cite{li2020early, zhou2020clinical} respectively.

Next, for an individual under $S$ state at time $t$, the probability of becoming presymptomatic state $P$ at $t+1$ is:
\begin{equation}
P_i^1(t+1) =S_i(t)\cdot(1-p)\cdot(1-\prod_{j\in \partial i}(1-\mathcal{F}(t,j,\beta))
\end{equation}
Accordingly, we can calculate the probability that the state of $i$ become $A$ at $t+1$ as:
\begin{equation}
A_i^1(t+1) =S_i(t)\cdot p\cdot(1-\prod_{j\in \partial i}(1-\mathcal{F}(t,j,\beta))
\end{equation}
In the third case where a node $i$ is in the infected state (i.e. $E$, $I$ or $A$, {$d \geq 1$}) on day $t$, the transition probabilities that they will stay in the same state on day $t+1$ are:
\begin{equation}
\begin{aligned}
&P^{d+1}_i(t+1) = P^d_i(t)\cdot(\frac{1-F_P(d)}{1-F_P(d-1)})\\
&I^{d+1}_i(t+1) = I^d_i(t)\cdot(\frac{1-F_{I}(d)}{1-F_{I}(d-1)})\\
&A^{d+1}_i(t+1) = A^d_i(t)\cdot(\frac{1-F_{A}(d)}{1-F_{A}(d-1)})
\end{aligned}
\end{equation}
where $F_P(d)=\int_{-\infty}^{d}{f_P(t)dt}$, $ F_{I}(d)=\int_{-\infty}^{d}f_I(t)dt$, $ F_{A}(d)$ $=\int_{-\infty}^{d}f_A(t)dt$ are the cumulative distribution functions of duration length $d$ for $P, I, A$ states, respectively. For mathematical convince, we simply set $F_P(0)=F_{I}(0)=F_{A}(0)=0$. The fourth case is that individual in the presymptomatic state $P$  turns into the symptomatic infectious state $I$ at the next day, and can be described with the following transition probability:
\begin{equation}
I^1_i  = \sum_{d=1}^{\infty} P^d_i\cdot\frac{F_P(d)-F_P(d-1)}{1-F_P(d-1)}
\end{equation}		
In the fifth case, an individual in the state $I$ or the state $A$ has a certain probability of being recovered i.e, turning into the  $R$ state on the next day. From the above equation, we obtain the probability that the individual $i$ is in the state of $R$ at the time $t+1$ is:
\begin{equation}
\begin{aligned}
R_i(t+1)  = &R_i(t) + \sum_{d=1}^{\infty} A_i^{d}\cdot\frac{F_{A}(d)-F_{A}(d-1)}{1-F_{A}(d-1)}\\ +&\sum_{d=1}^{\infty} I_i^{d}\cdot\frac{F_{I}(d)-F_{I}(d-1)}{1-F_{I}(d-1)}
\end{aligned}
\end{equation}

        %The vertex infection state probability transfer equation consists of the following five cases.
{To validate our mathematical framework, we test it on a real contact-tracing network in the \textit{Infectious Stay Away} exhibition\cite{isella2011s}~(ISA network, see Sec. SI 1 for data detailed description) with 410 individuals and average degree $\langle k \rangle$ of 13~(more experiments on another social network are illustrated in Sec. SI 5). } {We simulate the spreading with the empirically observed parameters on COVID-19 spreading mechanisms \cite{li2020early}~(see Methods for the simulation details). From repeated simulations, we then obtain the probability of every possible state of a node, and compare this baseline with the theoretical results from Eqns.~3-9.
%From repeated simulations, we then obtain the probabilities of each states that a node can take, and compare this baseline with the theoretical results from Eqns.~3-10.
Here we set the dimension of $Z$ to 77 according to the empirical temporal distributions of the infected states \cite{li2020early,zhou2020clinical}~(see Method for detail).
%The dimension of the state vector $Z$ corresponds to the total number of sub states possible during the various disease progression paths, i.e. the number of days that an individual can be in each of the 3 different infected states. From the empirical temporal distributions of the infected states [15,27] (Tab.~1), we use 3 standard deviations [8] as cut off on the max number of days in states $P,I,A$, which are $20,20,35$ days, yielding a dimension of 77 for $Z$. ($S$ and $R$ states have no sub states).
From Fig.~2a and b, we can see that our theoretical result on the temporal evolutions of the disease in the whole network is well validated by the simulations. These show that our transition probability framework is accurate in producing the real spreading dynamics.}
%The slight differences between the two {may} attributed to the short loops in the network~\cite{mezard2009information}.
%Additionally, the simulations validate our theoretical results on individual nodes in the middle of the spreading process in Fig.~2b~($T=10$).
%as well as the temporal evolution of a random node (Fig.~2c-f).

\begin{table*}[h]
	\centering
	\caption{COVID-19's clinical parameters and infectious characteristics used in this work.}
	\begin{tabular}{p{1.8cm}|p{3.3cm}|p{4.3cm}|p{4.3cm}}
		\hline
		\hline
		Parameter & Meaning& value&Origin  \\
		\hline
		$R_0$ & basic reproduction number & 3.50& Average from 10 researches~\cite{tang2020estimation,riou2020pattern,zhao2020preliminary,imaireport,cao2020estimating,wu2020nowcasting,shen2020modelling,liu2020transmission,read2020novel,majumder2020early} \\
		\hline
		$p$& fraction of asymptomatic infections & 15\% & minimal value from 5 researches\cite{nishiura2020estimation,byambasuren2020estimating,nishiura2020estimation,yu2020covid,mizumoto2020estimating}\\
		\hline
		$f_P(d)$ & Distribution of during length of Presymptomatic state & Logarithmic normal distribution with $\mu_P=0.62$ and $\sigma_P=0.64$& fitted value from clinical data of \cite{li2020early}\\
		\hline
		$f_{I}(d)$ &  Distribution of during length of Symptomatic state &Normal distribution with $\mu_{I}=8.8$ and $\sigma_{I}=3.88$ &fitted value from clinical data of \cite{li2020early}\\
		\hline
		$f_{A}(d)$& Distribution of during length of Asymptomatic state & Normal distribution with $\mu_{A}=20.0$ and $\sigma_{A}=5.0$ & $\mu_{A}$ is estimated from clinical data of~\cite{zhou2020clinical}\\
		\hline
		\hline
	\end{tabular}
	\label{tab1}
\end{table*}

{Now we extend the proposed transition probability equations to identify nodes with high risk of being asymptomatic, assuming the infection history on symptomatic nodes is already known.
The underlying principle is to update every node's state by incorporating the information of known infection into Eqns.(2-9) in the subsequent days, and then deduce the infection probability $C_i(T)$ for each node $i$ in the network~(see the details in the Method).
The nodes with higher $C_i(T)$ are identified as having high risk of being infected at day $T$.} We test the effectiveness by applying it on two sets of real COVID-19 spreading data on the contact-tracing network in Singapore~\cite{againstcovid19.com} (see Fig.~2c and d). {The details of network is provided in Method and in Sec. SI 1.} We find that the ranking our $C_i(T)$ values are highly correlated with the date of infection $t$ of nodes (Fig.~2e and f), meaning nodes with higher infection probabilities indeed have higher risk of being infected in the real COVID-19 spreading data.

The Singapore empirical  datasets have the constraint of merely including the symptomatic individuals' identities in the network.
Therefore, to further evaluate our method, we simulate a realistic COVID-19 spreading process on the ISA network for $T$ days to obtain the detailed infection history of every node in the network, such that the exact infection history on the asymptomatic nodes can be obtained. Assuming only the symptomatic nodes with state $I$ are observed, we use our above method to identify those infected individuals among the rest of the nodes. Specifically, we select the nodes with the highest $C_i(T)$ values as the mostly likely infected nodes. To our best knowledge, there is few prior works for estimating asymptomatic nodes in the network. Therefore, we also design several screening baselines based on the popular graph neural networks methods including Node2Vec \cite{grover2016node2vec}, graph convolutional network ~\cite{kipf2016semi} and graph attention networks \cite{velivckovic2017graph} to further compare our results. ({Detailed methodologies for those methods in Sec. SI 3}).

The simulations results  show that our transition probabilistic method (i.e. static screening) significantly outperforms the other methods in terms of the accuracy and recall on the local network where one can only observe the nearest neighbors of the known nodes in states $I$ ({see SI Figure 1}).
Such advantage is still evident when we consider the alternative scenario that one can observe the full network structure~\cite{drew2020rapid,ferretti2020quantifying}~({see SI Figure 3}), and the intermediate scenario when only nearest and second nearest neighbors are known in the network ({see SI Figure 2}).
In a more realistic setting, the screening of the contact tracing network happens continuously in time. Here one can update the set of known infected nodes after every screening, and subsequently update the infected risk for the rest of the network from time to time. Therefore, we develop a dynamic screening method by updating the evaluation of $C_i(t)$ every time a new infected node is found through selective screening of the network~(see the details in the Method Section). This dynamic screening method outperforms~ (see Fig.~3a-d) other screening methods and even our previous static screening method (see Fig.~3d inset), implying that such dynamic screening method is highly effective in identifying infected nodes by screening less people.

%*******HU  In a more realistic setting, the screening of the contact tracing network happens continuously in time. Here one can update the set of known infected nodes after every screening, and subsequently update the infected risk for the rest of the network from time to time. Therefore, we develop a dynamic screening method by updating the evaluation of $C_i(t)$ every time a new infected node is found through selective screening of the network~(see the details in the Method Section).  The simulations results show that our transition probabilistic method significantly outperforms~ (see Fig.~3abcd) the other methods in terms of the accuracy and recall, implying that such dynamic screening method is highly effective in identifying infected nodes by screening less people. Such advantage is still evident for static screening method on the local network where one can only observe the nearest neighbors of the known nodes in states $I$ ({see SI Figure 1}), on the alternative scenario that one can observe the full network structure~\cite{drew2020rapid,ferretti2020quantifying}~({see SI Figure 3}), and the intermediate scenario when only nearest and second nearest neighbors are known in the network ({see SI Figure 2}).

Very often, the  contact tracing network collected through either manual survey or digital tracking is at best incomplete, such that it is important to have a screening method that is still robust when there is missing information on the network structure. To test such robustness of our method, we randomly remove up to 80\% of the edges in the ISA network, and test the accuracy of the method based on the remaining network~{(see the results on another network in SI Figures 14-19)}.
We find that the dynamic screening method on the various scenarios can still reliably identify the infected nodes in terms of accuracy and recall rate, as shown in Fig.~3e-h~(see the robustness result on the static screening method in SI Figures 14-17).
%We find that the static screening method on the various scenarios can still reliably identify the infected nodes with consistent accuracy and recall rate ({see SI Figures 14-17 }).
%Intriguingly, our method on the various scenarios can still reliably estimate the infected nodes with consistent accuracy and recall rate (Fig.~5ac). Such consistency against various level of missing link information is shown in Fig.~5bd, as accuracy and recall rate are hardly affected even with only 40\% of the links remaining.
%The robustness result on the dynamic screening method is shown in Fig.~3e-h. It is worth noting in Fig.~3f that, under extremely high level of missing links, the unknown nodes that are connected with several known infected nodes have a high chance of remaining in the network, resulting in a high accuracy, but the recall rate is very low, meaning that a large portion of unknown infected nodes are not identified.

Lastly, we study the effectiveness of our method in containing the overall spread of COVID-19. In the widespread of COVID-19, limited resource on screening constrains the number of individuals the government can screen in a given day. Hence, targeted screening and mitigation can have significant impact on `flattening the curve' of daily infected cases. To study such effect, we again simulate the COVID-19 spreading on the ISA network \cite{isella2011s}, and start screening/testing from day 10 using our method (i.e. `neighbor containment'). Each day 2\% (4\% in SI Figure 20) of the whole network are tested for the disease, and the positive ones are immediately quarantined, corresponding a transition to the state $R$~(see details of the containment strategy in Sec. SI 4). As shown in  Fig.~4a-e, our method is highly effective in suppressing the daily infection cases and total infection cases, outperforming both the baseline strategy of only quarantining the infected ones (labeled as `infection containment') and the strategy randomly screening $2\%N$ among the neighbors of the known infections (labelled as `neighbor containment'), where $N$ is the network size. In addition, we find that even with up to 80\% missing links, our method is still robust enough to effectively suppress the spreading, close to that of knowing the full network structure. It shows that our method is expected to be highly effective in containing COVID-19 spread in practice.

\section{Conclusion}
In this paper, based on the transmission rule of COVID-19 and the underlying physical spreading equations, we for the first time studied the estimation of asymptomatic infections in the contact-tracing network, which is a current major concern in the prevention and containment of COVID-19 worldwide. We provided a complete computational framework of inferring latent infection on contact network. Based on this, we proposed a feasible method for optimal detection of latent infection in combination with nodal transmission ability in the network. We show that the COVID-19 transmission can be broken in a timely and efficient manner by the proposed method, which outperforms the direct contact screening, a typical method widely used in China. In addition, our simulation on a real contact network demonstrated that, this method is  robust even with incomplete network information, demonstrating its effectiveness in practical scenarios. We believe that the theory and the corresponding methods in identifying COVID-19 hidden spreaders are of great practical significance. In principle, it provides policymakers and front-line workers in COVID-19 with important and effective guidance and tools that could be deployed swiftly to fight COVID-19, and  save billions of people around the world who are still suffering as the epidemic continues to spread throughout the world.

\section*{Method}
\textbf{Singapore COVID-19 datasets.} The data was collected by the Singapore government~\cite{againstcovid19.com}, and contains comprehensive records on the dates of showing symptoms and confirming the disease, as well as their contact networks. We pick the infected nodes from the first two different time points, and set $T$ to January 26, 2020 and February 19, 2020 for Singapore A and Singapore B, respectively. Then based on the known infection history of the nodes, our transition probability method estimates the infection probabilities  $C_i(T)$ of every other node in the networks.

\textbf{The dimension of the state vector $\mathbf{Z}_i(t)$.} The dimension of the state vector $\mathbf{Z}_i(t)$ corresponds to the total number of sub states possible during the various disease progression paths, i.e., the number of days that an individual can be in each of the 3 different infected states. From the empirical temporal distributions of the infected states~\cite{li2020early,zhou2020clinical}~(Tab.~1), we use 3 standard deviations~\cite{wheeler1992understanding} as cut off on the max number of days in states $P,I,A$, which are $20,20,35$ days, yielding a dimension of 77 for $Z$. ($S$ and $R$ states have no sub states).

\textbf{COVID-19 spreading simulation.} At the starting time $T=0$, we select the 3 nodes with the largest degree in the network as initial infected nodes, whose infected states are determined as either Asymptomatic or Presymptomatic according to the parameter $p$ of Tab. 1. Then we apply the empirically observed parameters on COVID-19 spreading mechanisms \cite{li2020early} including reproductive number $R_0=3.50$ and asymptotic infection ratio $p=15\%$ on our equations to simulate the spreading. The set of values are listed in Table.~1(see Sec. SI 1 for the detail description of parameters and Sec. SI 3 for the discussion of the parameter sensitivity). Each simulation corresponds to one realization of the actual spreading based on the realistic dynamics, and the actual states of each node at every time step can be captured. More details of the simulation of COVID-19 are provided in Sec. SI 2.

\textbf{Identifying infection probability $C_i(T)$.} The goal is to identify nodes with high risk of being asymptomatic with infection history on known symptomatic nodes, and we extend our transition probability equations to study this problem. At a certain time $T$, given the set $I$ of infected individuals, the first day of infection $s_j$ and the day of recovery $r_j$ for each individual $j\in I$, we aim to develop a method from Eqns.~2-10 to deduce the infection probability $C_i(T)$ for each node $i$ in the network. Note that the day of recovery can also be the day of death or quarantine.
%since effectively the node has been removed from the network spreading process.
The initial condition at $t=0$ is that every node in the network is in susceptible state, i.e. ${\bf Z}_i(0)=\{1,0,\cdots,0\}$. The day of first infection in the network is set to 1, i.e. $\min_{j\in I}s_j=1$, and we update every node's state in the subsequent days depending on whether their infection history is known at time $T$. For the known nodes $j\in I$, we artificially assign their infection states according to the known information, meaning that $j$ is assigned state $S$ when $t<s_j$,  state $R$ when $t>r_j$, and infectious state when $s_j<t<r_j$. For the other nodes, we evaluate their state vector $Z_i(t)$ at every time step $t$ according to the transition probabilities in Eqns.~2-9, until the final day $T$, such that their probabilities $C_i(T)$ of being infected can be evaluated from $Z_i(T)$.

\textbf{Dynamic screening method.} Every time we screen only  node $k$ that is of highest risk according to the algorithm; if node $k$ is COVID-19 positive, it is added to the known infected nodes set $I$, and its neighbors are added to the unknown set, and we repeat the transition probability calculations according to Eqns.~2-10 from time $0<t\le T$; if $k$ is negative, its probability state vector is set to be ${\bf Z}_k(t)=\{1,0,\cdots,0\}$ in the calculation of Eqns.~2-10. Next, the revised estimations of infection probabilities for each unknown node from Eqns.~2-10 tells us which node is the most risky and to be tested.

\bibliographystyle{unsrt}
\bibliography{COVID19-Find}
\clearpage
\begin{figure*}
\includegraphics[width=1\textwidth]{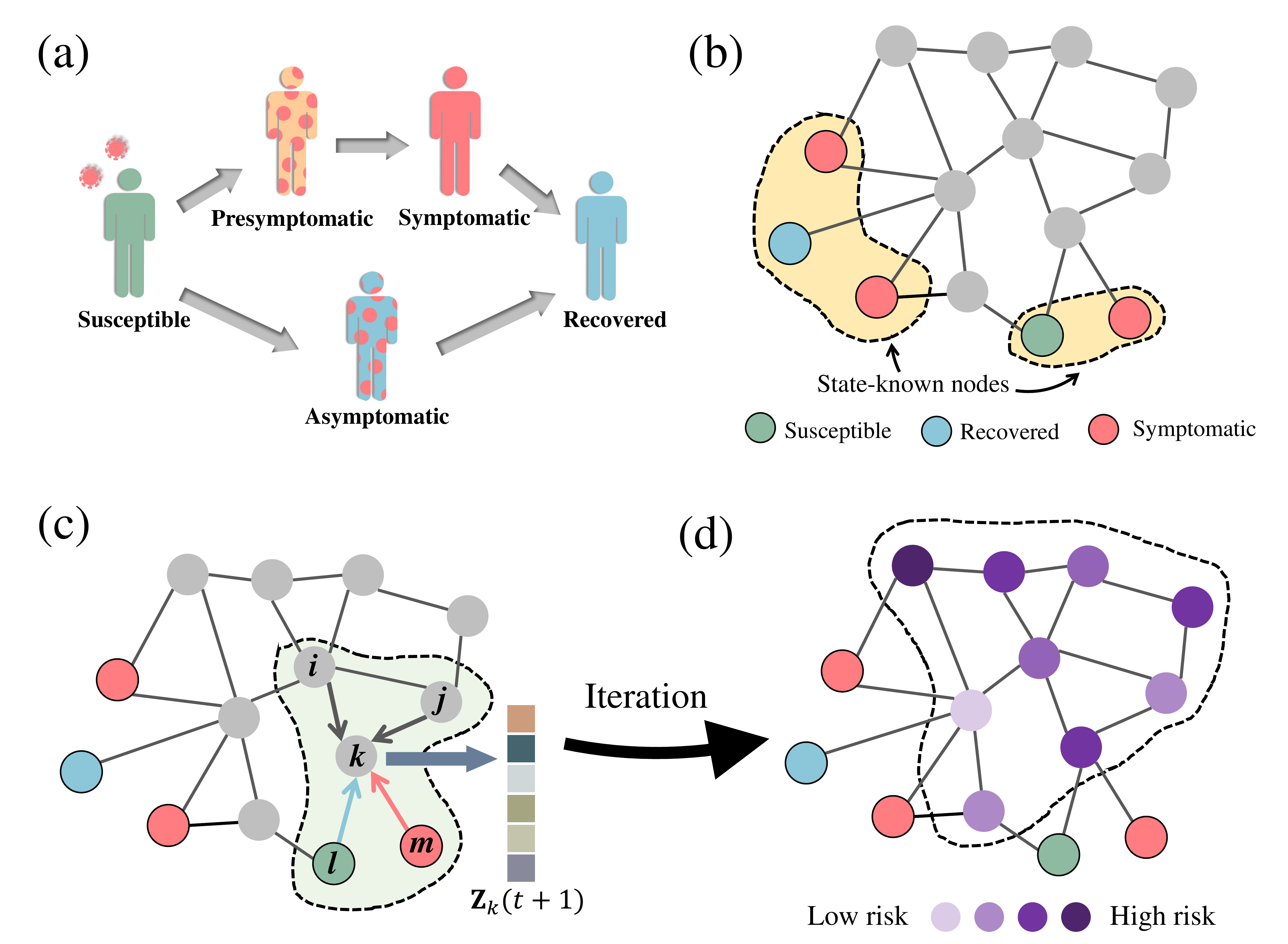}
\caption{\textbf{Identifying Asymptomatic and Presymptomatic COVID-19 Infections on Contact-tracing Networks}. (a) The COVID-19 state transition of an individual. A susceptible will become either asymptomatic or presymptomatic after being infected. An asymptomatic patient will further turns to the symptomatic state after an incubation period, while an presymptomatic patient will never exhibit any symptoms until the recover.
(b) Illustration of asymptotic/presymptomatic node identification problem. In a contact-tracing network, only a small fraction of the nodes' states are known~(marked with color), while the hidden asymptomatic/presymptomatic individuals within the population~(marked with grey) are potential spreaders. Our purpose is to find asymptomatic/presymptomatic infected individuals in the population  using the contact-tracing network and the information of known confirmed cases. (c) Diagram of the proposed method. The state transition of an unknown node $k$ is modeled as a Markov process, i.e., a vector $\mathbf{Z}_k$ where the elements represent the probabilities of different infection stages in (a). The specific value of the vector $\mathbf{Z}_k(t+1)$ at $t+1$ is determined by the infection status of known nodes at $t$ and the structure of the contact network.
(d) After iterations over the whole network, each unknown node will be assigned with an infection indicator according to the eventual values of its state vector $\mathbf{Z}_k$, which represents the risk of being infected.}
\label{Fig1}
\end{figure*}
\begin{figure*}
\centering
\includegraphics[width=1\textwidth]{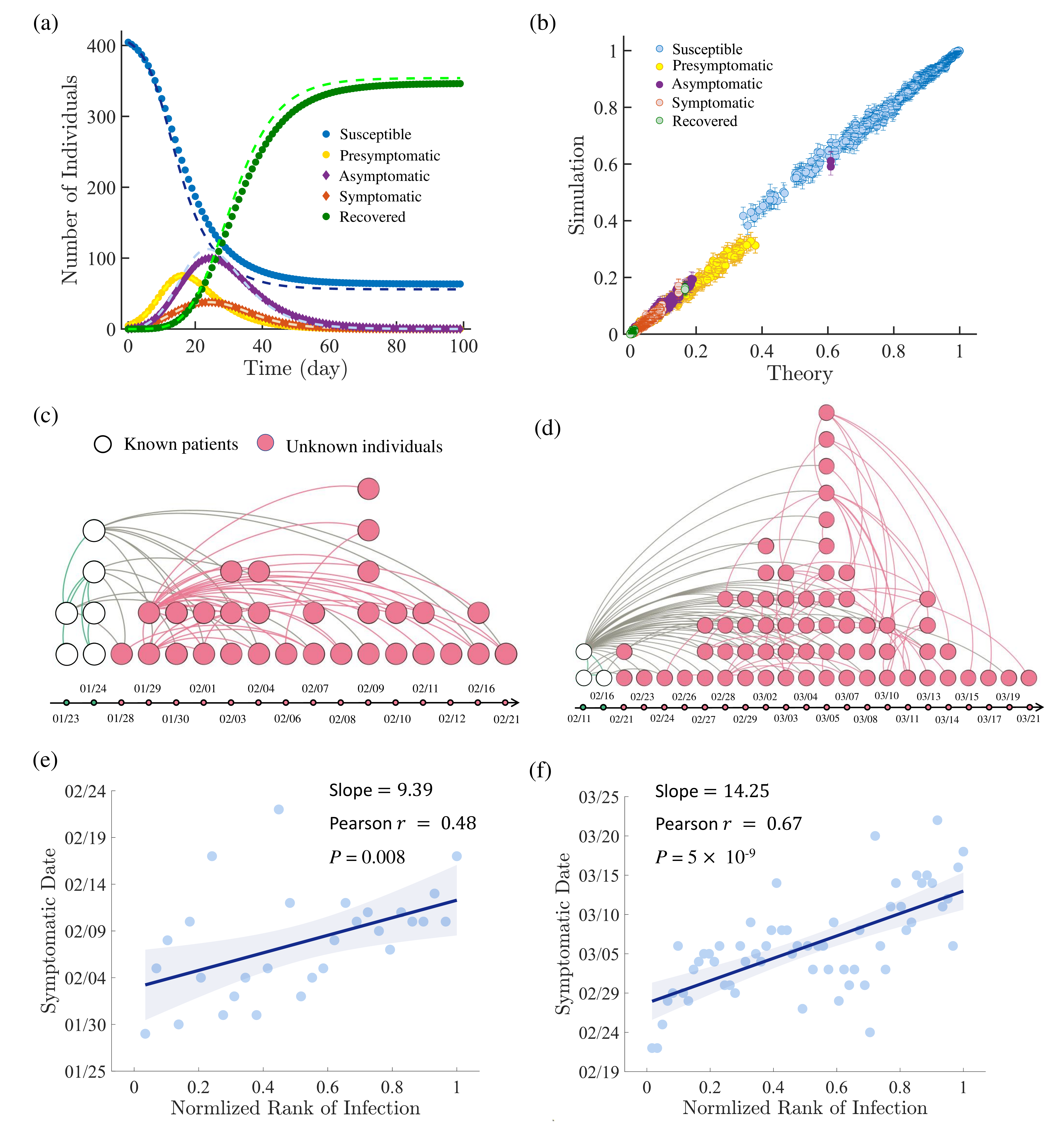}
\end{figure*}
\begin{figure*}
\caption{\textbf{Empirical and Simulated Validation of the Proposed Model.}
(a) The number of people in each of the 5 states in the simulation process of COVID-19 spreading on the ISA network. At $T=0$, we select three nodes with the maximum degree in the ISA network as the initial infected spreaders. Dash lines represent the theory values calculated by Eqns.(6-10). Dots represent the average value of 1000 simulations.
(b) The theoretical probability vs. numerical frequency of each individual being in various states on $T= 10$ days. Each dot corresponds to a certain state of a node in the ISA network while the errorbar is the 95\% confidence interval obtained by the bootstrapping method~\cite{diciccio1996bootstrap}.
(c) The topological structure of a real COVID-19 spreading network in Singapore~(Singapore A), where dots are patients and curves are contacts between patients (see Sec. SI 1 for the description of the network). The points on the timeline indicate the date of the patient's presence. (d) The topological structure of another network of Singapore~(Singapore B).
(e) Relationship between the individual symptomatic time and the estimated infection probability in Singapore A. The network has a total of $T=30$, from January 23, 2020 to February 21, 2020. Here we utilize the information of infections from the first two different time points as known set to infer the rest nodes' states in the network by Eqns.~(1-9).
%Using the obtained state vector of each unknown node, we rank them according to the infection probability $C_i(t)$ and compare with it's real symptomatic time. Each dot corresponds to a single infection.
Using the obtained state vector of each unknown node, we rank them according to the infection probability and compare with it's real symptomatic time. Since all patients are symptomatic in the dataset, the rank is based on $C_i(t)-A_i(t)$.
{The line denotes the linear fitted result and the shaded area denotes the 95\% confidence interval.}
(f) Similar to (e), the value of the rank of infection probability vs. symptomatic time in Singapore B. The network has a total of $T=40$, from February 11, 2020 to March 21, 2020. We use infections who got infected the first two different time points for training.
}
\label{Fig2}
\end{figure*}

\begin{figure*}
\centering
\includegraphics[width=0.8\textwidth]{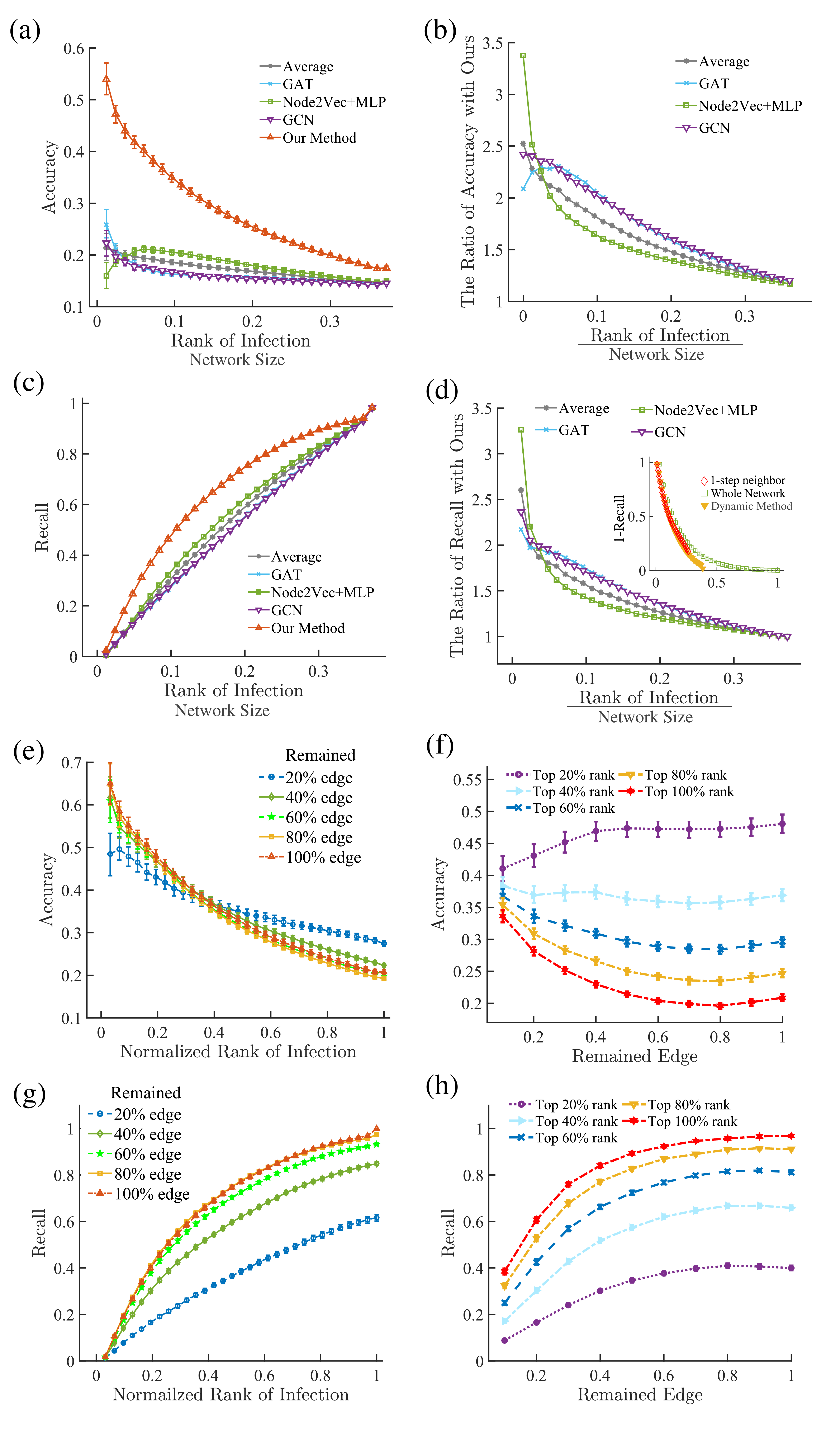}
\label{Fig3}
\end{figure*}
\newpage
\begin{figure*}
\caption{{\textbf{Screening Performance Assessment.} (a-d) Performance of the dynamic screening method on the ISA network compared with other machine-learning baselines~(see details in Sec. SI 3). (a) The accuracy vs. rank of infection~(divided by the network size). The accuracy is defined as the proportion of non-Susceptible individuals in the ranking list. {Since all nodes we screen at $T$ do not have symptoms, here we use the value of $C_i(t)-I_i(t)$ to rank these nodes.} (b) Similar to (a), the relative accuracy of the machine-learning-based algorithms, which is the ratio between the accuracy of our proposed algorithm and other algorithms. (c) The relationship between the rank of infection and the recall rate (i.e., the proportion of successfully identified non-Susceptible individuals to those in the whole network. (d) Similar to (c), the relative recall rate of the machine-learning-based algorithms. (inset) Recall rate of the static algorithms on the 1-step neighbor subnetwork and on the whole network~({see SI Figures 14-17}), compared with the dynamic algorithm. (e-h) Performance of the dynamic screening method with incomplete network information. We randomly remove a fraction of links in the ISA network~(see the result of another social network in Sec. SI 5) and then employ the proposed screening schemes on the remaining network. (e) The relationship between the accuracy and the ranking value of the infection probability with different proportions of the removed edges. Here we normalize the ranking value to range [0,1] by dividing it with the total number of individuals who have been screened. (f) Accuracy of the dynamic screening method vs. the proportion of the removed edges by measuring the infection rank with different proportions. {For example, top 20\% rank means the 20\% nodes with the highest infection possibility which is equal to 0.2 of norlized rank of infection in (e).} (g) The dependencies of recall rate of the dynamic screening method on the infection rank. (h) Recall vs. the remaining edges.}}
\end{figure*}

\begin{figure*}
\centering
\includegraphics[width=1\textwidth]{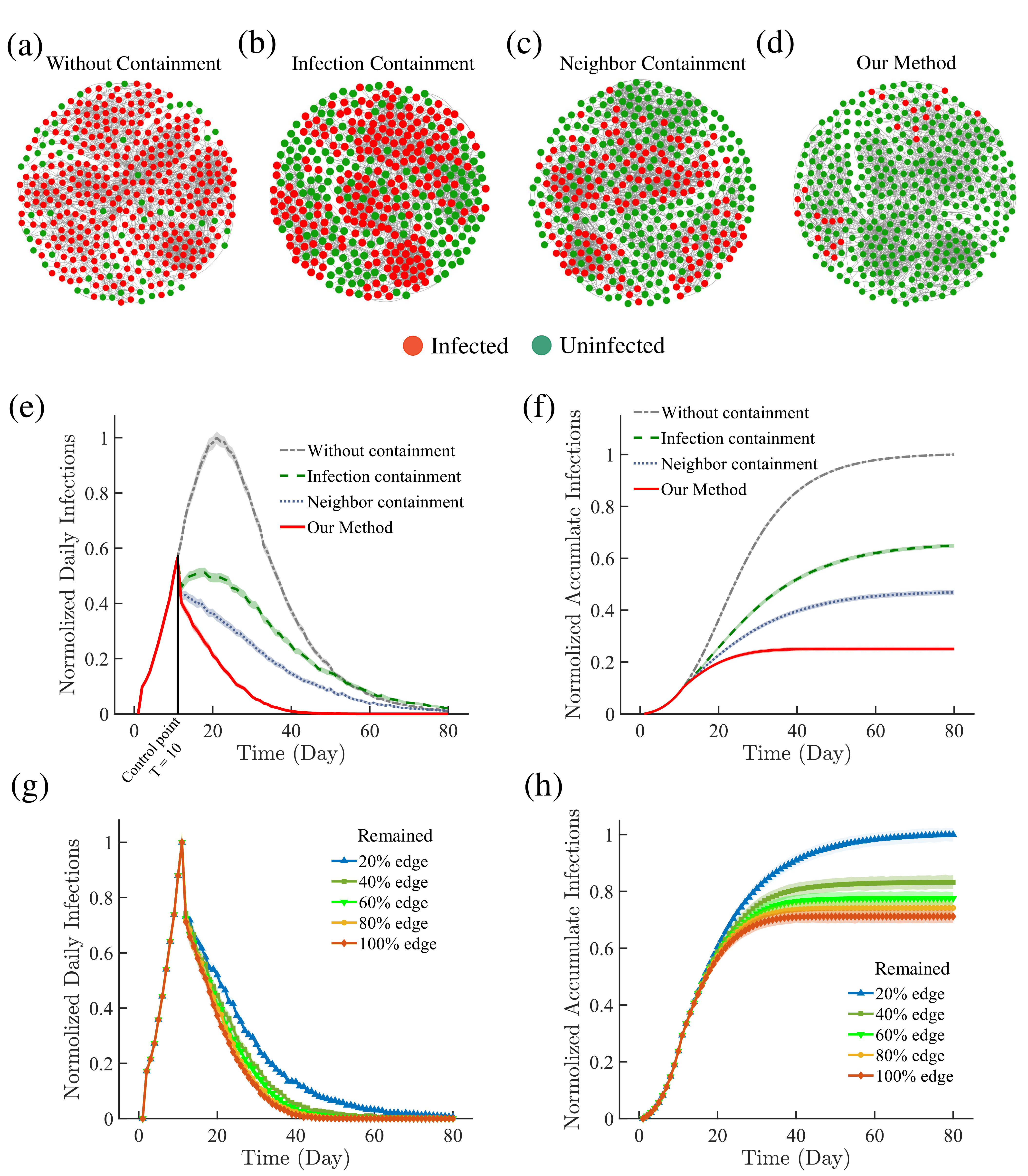}
\caption{\textbf{Containment Effectiveness of the Proposed Scheme on Simulated COVID-19 Spreading.} From $T=10$ of COVID-19 spreading simulation on the ISA network, we conduct different approaches separately to contain the pandemic, including the proposed method (i.e. Dynamic Containment), the Infection Containment and the Neighbor Containment (see Sec. SI 4 for details). For the Neighbor Containment and the proposed method, we select a total of $2\%N$ of individuals to screen and then quarantine those tested positive at each time step. (a) Visualization of infected population without conducting any contain scheme and (b-d) the 3 different control schemes). (e) The number of daily new infections in different control schemes. The value of each curve is divided by the highest point at that time to normalize to [0,1]. (f) The cumulative number of infections corresponding to (e). (g-h) The performance of our control scheme under an incomplete network where a fraction of links are randomly removed in the detection process.}
\label{Fig6}
\end{figure*}

\end{document}